\documentstyle[aas2pp4]{article}

\def\simless{\mathbin{\lower 3pt\hbox
     {$\rlap{\raise 5pt\hbox{$\char'074$}}\mathchar"7218$}}}   
\def\simmore{\mathbin{\lower 3pt\hbox
     {$\rlap{\raise 5pt\hbox{$\char'076$}}\mathchar"7218$}}}   

\received{}
\accepted{}
\slugcomment{Submitted to Ap.J.Letters., February 1999}

\lefthead{M.~M\'endez et al.}
\righthead{Kilohertz QPOs in 4U 1728--34}

\begin{document}

\title{Precise Measurements of the Kilohertz Quasi-Periodic
Oscillations in 4U 1728--34}

\author{Mariano M\'endez\altaffilmark{1,2},
        Michiel van der Klis\altaffilmark{1}
}

\altaffiltext{1}{Astronomical Institute ``Anton Pannekoek'',
       University of Amsterdam and Center for High-Energy Astrophysics,
       Kruislaan 403, NL-1098 SJ Amsterdam, the Netherlands}

\altaffiltext{2}{Facultad de Ciencias Astron\'omicas y Geof\'{\i}sicas, 
       Universidad Nacional de La Plata, Paseo del Bosque S/N, 
       1900 La Plata, Argentina}

\begin{abstract}

We have analyzed seventeen observations of the low-mass X-ray binary and
atoll source 4U 1728--34, carried out by the {\em Rossi X-ray Timing
Explorer} in 1996 and 1997.  We obtain precise measurements of the
frequencies of the two simultaneous kilohertz quasi-periodic
oscillations (kHz QPOs) in this source.  We show that the frequency
separation between the two QPO, $\Delta \nu$, is {\em always}
significantly smaller than the frequency of the nearly-coherent
oscillations seen in this source during X-ray bursts, even at the lowest
inferred mass accretion rate, when $\Delta \nu$ seems to reach its
maximum value.  We also find that $\Delta \nu$ decreases significantly,
from $349.3 \pm 1.7$ Hz to $278.7 \pm 11.6$ Hz, as the frequency of the
lower frequency kHz QPO increases from 615 to 895 Hz.  This is the first
time that variations of the kHz QPO peak separation are measured in a
source which shows nearly-coherent oscillations during bursts.

\end{abstract}


\section{Introduction}

Quasi-periodic oscillations at frequencies near 1 kilohertz (kHz QPOs)
are now known to be a common phenomenon among low mass X-ray binaries
(LMXBs).  Since the beginning of the {\em Rossi X-ray Timing Explorer
(RXTE)} mission kHz QPOs have been discovered in the persistent flux of
nineteen LMXBs (see van der Klis \cite{vanderklis98} for a review).  In
almost all cases the power density spectra of these sources show twin
kHz peaks that, as a function of time, gradually move up and down in
frequency, typically over a range of several hundred Hz.

The dependence of the QPO frequencies upon X-ray flux is complex.  One
example is 4U 1608--52 (M\'endez et al.  \cite{mendez99}):  While on
time scales of hours frequency and X-ray flux are well correlated, at
epochs separated by days to weeks the QPOs span the same range of
frequencies even if the average flux is different by a factor of 3 or
more (see also Ford et al.  \cite{ford97a}; Zhang et al.
\cite{zhangAqlx1}).  Ford et al.  (\cite{ford97a}) have shown that in 4U
0614+09 the QPO frequency is better correlated to the flux of the soft
(``blackbody'') component in its X-ray spectrum than to the total X-ray
flux, while Kaaret et al.  (\cite{kaaret98}) found, both for 4U 0614+09
and 4U 1608--52, a good correlation between the QPO frequency and the
photon index of the power-law component in the X-ray spectrum.  M\'endez
et al.  (\cite{mendez99}) showed that QPO frequency is very well
correlated to the position of the source on the color-color diagram, and
they concluded that most likely these frequency changes are driven by
changes in $\dot M$ through the innermost part of the accretion disk,
and that there is no one-to-one relationship between the observed X-ray
flux and this mass accretion rate.

Initially, data from various sources suggested that the separation
$\Delta \nu$ between the twin kHz peaks remained constant even as the
peaks moved up and down in frequency (e.g., Strohmayer et al.
\cite{strohmayer96}).  In some sources a third, nearly-coherent,
oscillation is detected during type-I X-ray bursts, at a frequency close
to $\Delta \nu$, or twice that value (see Strohmayer, Swank, \& Zhang
\cite{ssz98} for a review).  These two results suggested that a
beat-frequency mechanism was at work (Strohmayer et al.
\cite{strohmayer96}; Miller, Lamb, \& Psaltis \cite{miller98}), with the
third peak being close to the neutron star spin frequency or twice that.

It is now clear that $\Delta \nu$, which in the beat-frequency
interpretation is thought to be equal to the spin frequency of the
neutron star, is not always constant.  In Sco X--1 (van der Klis et al.
\cite{vanderklisetal97a}), 4U 1608--52 (M\'endez et al.
\cite{mendez98c}), and 4U 1735--44 (Ford et al.  \cite{ford1735}),
$\Delta \nu$ slowly decreases as the frequencies of both QPOs increase.
In these sources no burst oscillations have been detected so far.  In 4U
1636--53 the frequency of the burst oscillations (or half this value)
does not exactly match the frequency difference between the kHz QPOs
(M\'endez, van der Klis, \& van Paradijs \cite{mendez98a}).  In this
case, however, no evidence was found of any significant change of
$\Delta \nu$.  So far, no measurements existed of burst oscillations and
a varying kHz peak separation in the same source.

In this paper we present such measurements for the first time.  We show
that in 4U 1728--34, the frequency separation between the two
simultaneous kHz QPOs is always lower than the frequency of the burst
oscillations, even at low kHz QPO frequencies, when the peak separation
appears to have reached its highest value.  We also find that $\Delta
\nu$ decreases significantly as the frequencies of the kHz QPOs
increase.  As in 4U 1608--52, the QPO frequencies in 4U 1728--34 are
well correlated to the position of the source on the color-color
diagram, despite them not being correlated to the X-ray flux.

\section{Observations}

{\em RXTE} observed 4U 1728--34 on ten occasions in 1996, between
February 15 and March 1, for a total of 199 ks (for a description of
these observations see Strohmayer et al.  \cite{strohmayer96},
\cite{strohmayer97}; Ford \& van der Klis \cite{ford&vdk98}), and seven
more times starting UTC 1996 May 3 13:56:00 for 3.9 ks, 1997 September
23 23:48:00 for 1.4 ks, September 24 09:13:00 for 22.9 ks, September 26
12:28:00 for 18.1 ks, September 27 09:17:00 for 18.3 ks, September 30
04:31:00 for 20.6 ks, and October 1 06:07:00 for 21.2 ks.

Here we only use data recorded by the Proportional Counter Array (PCA)
onboard {\em RXTE} (Bradt, Rothschild, \& Swank \cite{bradt93}).
Besides the two standard modes that are available to all {\em RXTE}/PCA
observations, data were collected using additional modes with high time,
and moderate energy resolution, covering the nominal $2-60$ keV energy
band of the {\em RXTE}/PCA.  Part of these data (the 1996 observations)
were analyzed previously (Strohmayer et al.  \cite{strohmayer96},
\cite{strohmayer97}).  Their analysis of the kHz QPOs was based on a
simple direct FFT method, while ours uses the ``shift-and-add'' method
(see below).

To estimate the source X-ray intensity and colors we measured the count
rates every 16 s in 5 energy bands, $2.0 - 3.5 - 6.4 - 9.7 - 16.0$ keV,
and $2.0 - 16.0$ keV, taking into account the gain changes applied to
the PCA in March and April 1996.  In a few of the observations one or
two of the five detectors of the PCA were switched off; we only used the
three detectors which were always on to calculate these count rates.  We
subtracted the background contribution in each band using the standard
PCA background model version 2.1b, and normalized the count rates to 5
detectors. For the timing analysis we used data from all active detectors
without energy selection.

\section{Results}

\subsection{Measurement of the frequencies of the QPOs}

We divided the high-time resolution data into segments of 64 s, and
calculated a power spectrum for each segment preserving the maximum
available Nyquist frequency (8192 or 65536 Hz, depending on the
observation).  We produced an average power spectrum for each
observation and searched it for QPOs at frequencies above 100 Hz.

In the rest of this subsection we only present the results of the
analysis of those 1482 data segments in observations where we detected
two simultaneous kHz QPOs in the power spectrum.  Here we concentrate on
the measurements of the frequency separation $\Delta \nu$ of these QPOs,
as other properties of the QPOs have already been reported in previous
papers (Strohmayer et al.  \cite{strohmayer96}; Ford \& van der Klis
\cite{ford&vdk98}; for the observations that we report here for the
first time, the other properties of the QPOs are similar to those
described in these two papers).

To measure $\Delta \nu$ we applied the `shift-and-add' technique
advanced by M\'endez et al.  (\cite{mendez98b}).  Measuring $\Delta \nu$
in an averaged (non-shifted) power spectrum can introduce a bias that
changes the peak separation from its true value because, in general, the
two QPOs vary in strength in a different manner as they move in
frequency (see for instance Wijnands et al.  \cite{wijnands1636}).  This
is no longer the case if the individual power spectra are shifted to the
frequency of one of the two peaks before they are averaged together.
(However, this method can be applied only if one of the QPOs is always
strong enough to measure its frequency in each of the individual power
spectra.)  A second advantage of this method is that longer stretches of
data can be used to measure $\Delta \nu$, and not just pieces of data
where the QPO frequencies are more or less constant.  This, in turn,
decreases the errors in the measurements of the frequency separation
(see below).

We measured the central frequency of the lower frequency QPO in each
segment of 64 s by fitting a Lorentzian to it (this QPO was well
detected in each segment).  We then grouped the data in 8 sets of 30,
58, 66, 193, 371, 262, 244, and 258 power spectra, such that the
frequency of the QPO did not vary by more than $\sim 50$ Hz within each
set (the first 2 sets contain data from 1996 only, whereas the last 2
sets contain data from 1997 only; segments 3 to 6 include data from both
epochs).  We then shifted the frequency scale of each spectrum to a
frame of reference where the position of this peak was constant in time,
and we finally averaged the aligned spectra in each set.

\begin{figure}[ht]
\plotfiddle{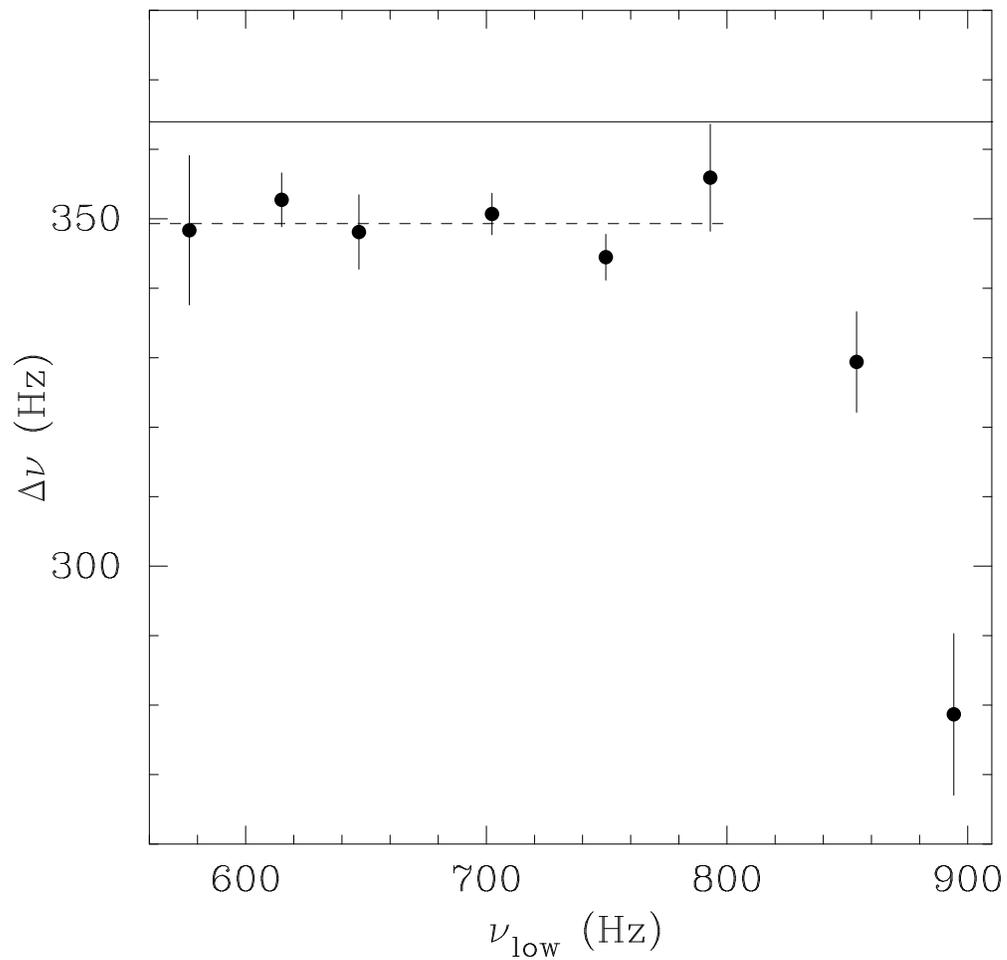}{220pt}{270}{70}{70}{-270}{250}
\vspace{6cm}
\caption{
The frequency separation between the upper and the lower QPO peak as a
function of the frequency of the lower QPO for 4U 1728--34 (circles).
The solid line indicates the frequency of the burst oscillations
at 363.95 Hz (Strohmayer et a. \cite{strohmayer96}, \cite{ssz98}).
The dashed line indicates the average value of $\Delta \nu$ (349.3 Hz)
for the first 6 points (see text).
\label{figdif}
}
\end{figure}

We fitted the resulting 8 average power spectra in the range $256 -
1500$ Hz using a function consisting of a constant, representing the
Poisson noise, and two Lorentzians, representing the QPOs.  The fits
were good, with reduced $\chi^{2} \leq 1.1$ for 1585 degrees of freedom,
and the significance of each peak was always $> 4\sigma$.  In Figure
\ref{figdif} we plot the frequency difference, $\Delta \nu$, between the
higher frequency and the lower frequency QPOs (hereafter the upper and
lower QPO, respectively) as a function of the centroid frequency of the
lower QPO, $\nu_{\rm low}$.  The solid line indicates the frequency of
the oscillations detected during the X-ray bursts, $\nu_{\rm burst} =
363.95$ Hz (Strohmayer et al.  \cite{strohmayer96}, \cite{ssz98}).
These results are entirely consistent with those of Strohmayer et al.
(\cite{strohmayer96}; Table 1), but our error bars, thanks to the use of
the ``shift-and-add'' technique, are much smaller than theirs.

Because of these small error bars, it is now apparent that $\Delta \nu$
is {\em always} smaller than $\nu_{\rm burst}$.  The figure shows that
$\Delta \nu$ is consistent with being constant at $\nu_{\rm low} < 800$
Hz, but decreases as $\nu_{\rm low}$ increases above 800 Hz.  The data
point near $\nu_{\rm low} = 900$ Hz is significantly ($6 \sigma$) below
the average of the other points.  If, conservatively, we discard both
points at $\nu_{\rm low} > 800$ Hz the average frequency separation
becomes $\Delta \nu = 349.3 \pm 1.7$ Hz (indicated by the dashed line in
Fig.  \ref{figdif}), still different from 363.95 Hz at an $8.6 \sigma$
level.  So, even in the range where $\Delta \nu$ appears to have reached
its saturation value, it is significantly below the frequency of the
burst oscillations.

\subsection{QPO frequency as a function of inferred $\dot M$}

\begin{figure}[ht]
\plotfiddle{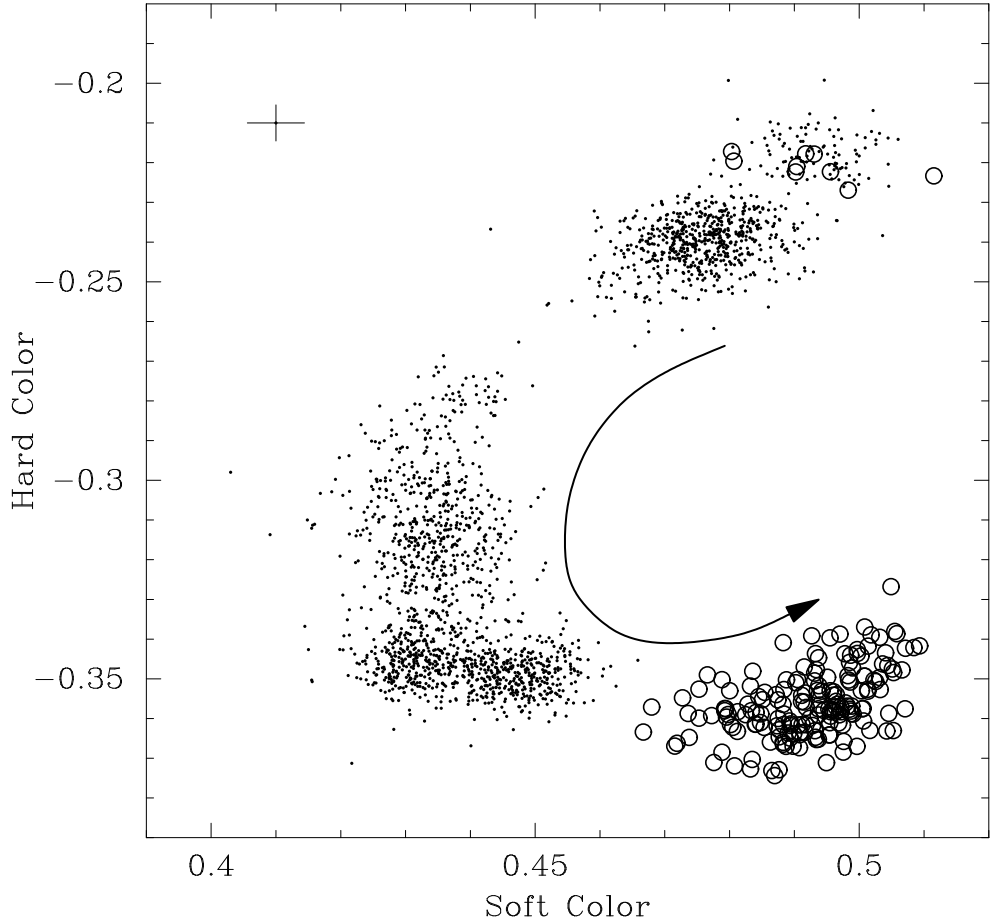}{220pt}{270}{70}{70}{-270}{250}
\vspace{6cm}
\caption{
Color-color diagram of 4U 1728--34.  The soft and hard colors are
defined as (the logarithm of) the ratio of count rates in the bands $3.5
- 6.4$ keV and $2.0 - 3.5$ keV, and $9.7 - 16.0$ keV and $6.4 - 9.7$
keV, respectively.  The contribution of the background has been
subtracted, but no dead-time correction was applied to the data (in this
case the dead-time effects on the colors are less than 1\,\%).  Each
point represents 128 s of data.  Typical error bars are shown. Dots and
open circles indicate segments with and without detectable kHz QPOs,
respectively. The arrow indicates the sense in which $\dot M$ is inferred
to increase.
\label{figcolor}
}
\end{figure}

In Figure \ref{figcolor} we show a color-color diagram of 4U 1728--34,
which is shown here for the first time to be typical of those of other
atoll sources (Hasinger \& van der Klis \cite{hasinger89}).  In general,
the count rate is observed to increase as indicated by the arrow on the
diagram, from the island state (at the upper right part of the diagram)
to the banana branch, but the relation between count rate and either of
the two colors is much less clean than that between colors.  Each point
in this diagram represents 128 s of data.  Dots and open circles
indicate the segments with and without detectable kHz QPOs,
respectively.

In Figure \ref{figrate} we show the frequency of the lower QPO,
$\nu_{\rm low}$, as a function of count rate for 4U 1728--34.  Each
point in this figure represents 128 s of data.  The branch in the lower
right part of the diagram corresponds to the data presented by
Strohmayer et al.  (\cite{strohmayer96}) for the lower kHz QPO.  This
plot can be compared to those in Aql X--1 (Zhang et al.
\cite{zhangAqlx1}) and 4U 1608--52 (M\'endez et al.  \cite{mendez99}).
As in those sources, for 4U 1728--34 it is also true that on time scales
greater than $\sim$ 1 day there is no unique relation between count rate
and $\nu_{\rm low}$.

As in the case of 4U 1608--52 (M\'endez et al.  \cite{mendez99}), QPO
frequencies correlate much better to the position of the source on the
color-color diagram than to count rate.  In Figure \ref{fig_sa} we show
$\nu_{\rm low}$ as a function of the position on the color-color diagram
as measured by the parameter $S_{\rm a}$ (curve length in the
color-color plane measured in the direction of the arrow in Fig.
\ref{figcolor}; see M\'endez et al.  \cite{mendez99} for the precise
definition of $S_{\rm a}$), for the same intervals shown in Figure
\ref{figrate} (black circles).  In this case $S_{\rm a}=1$ at colors
($0.48,-0.23$), and $S_{\rm a}=2$ at ($0.44,-0.34$).  The black squares
represent $\nu_{\rm upp}$, the frequency of the upper kHz QPO as a
function of $S_{\rm a}$ (we only include here segments where we could
measure $\nu_{\rm upp}$ directly).  The grey triangles are measurements
from segments in which we only detect one of the kHz QPOs; however, from
the location of each point in this diagram we can determine whether it
is the lower or the upper QPO.  While the frequency vs.  count rate
relation (Fig.  \ref{figcolor}) is complex, there seems to be a neat
relation between each QPO frequency and $S_{\rm a}$ (Fig.  \ref{fig_sa}).

\section{Discussion}

We have for the first time measured the variation of the kHz QPO peak
separation in a source which shows oscillations during X-ray bursts.
The result of this measurement is unexpected:  even at its saturation
level at low inferred $\dot M$, $\Delta \nu$ is still significantly
below the frequency of the burst oscillations.  Before we discuss this
further, we consider what our data imply about the relation between the
kHz QPO frequencies and $\dot M$.

\begin{figure}[ht]
\plotfiddle{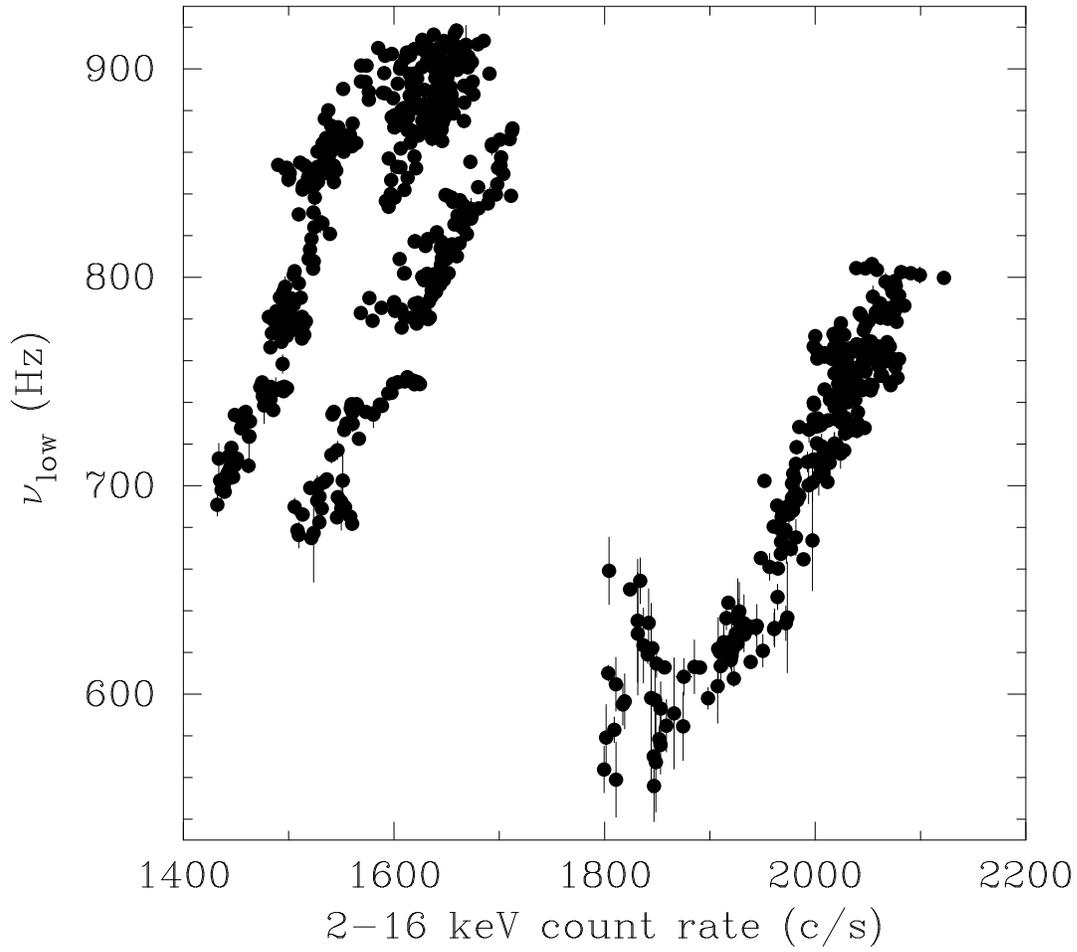}{220pt}{270}{70}{70}{-270}{250}
\vspace{6cm}
\caption{
Relation between the frequency of the lower kHz QPO and the $2 - 16$ keV
count rate.  The count rates have been corrected for background, and
normalized to 5 detectors.  Each point represents a 128-s segment.
We only includes data where both kHz QPOs were detected simultaneously,
so that we can unambiguously identify the lower kHz peak.
\label{figrate}
}

\end{figure}

\begin{figure}[ht]
\plotfiddle{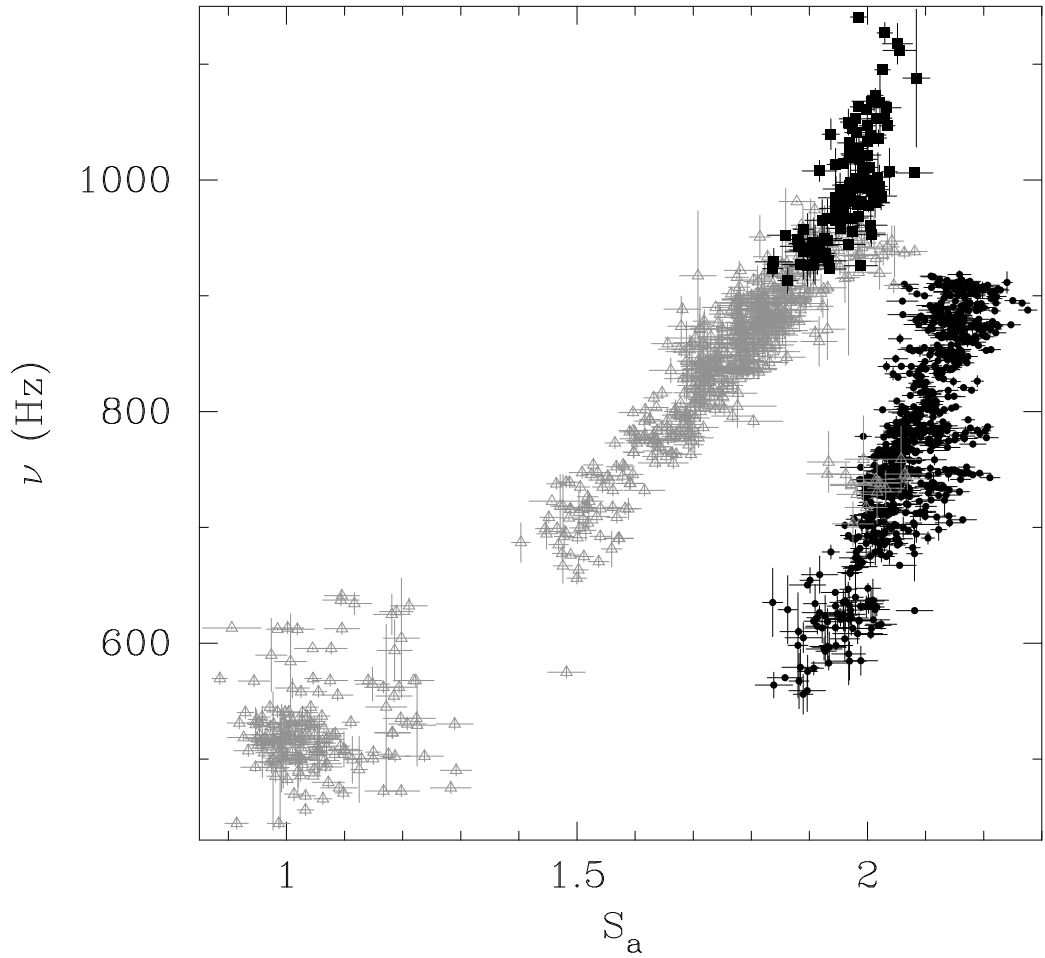}{220pt}{270}{70}{70}{-270}{250}
\vspace{6cm}
\caption{
Diagram of the frequencies of the upper and lower kHz QPOs vs.  the
position of the source on the color-color diagram, as measured by
$S_{\rm a}$ (see Fig.  \ref{figcolor}, and text).  Black circles and
squares represent the lower and the upper kHz peak, respectively.
Grey triangles represent segments where we only detected one of the kHz
QPOs in the power spectrum.
\label{fig_sa}
}
\end{figure}

Despite the complex relation between frequency and count rate (Fig.
\ref{figrate}), in 4U 1728--34 there is a unique relation between
frequency and position in the X-ray color-color diagram (Fig.
\ref{fig_sa}), similar to what is observed in 4U 1608--52 (see Fig.  2,
3, and 4 in M\'endez et al.  \cite{mendez99}).  As in 4U 1608--52, in 4U
1728--34 the frequencies of the QPOs increase with $S_{\rm a}$, as the
source moves from the island to the banana.  The presence of the QPOs
also correlates with the position in the color-color diagram.  This may
be a general feature.  In atoll sources it has been proposed that $\dot
M$ increases with $S_{\rm a}$ from the island to the banana (Hasinger \&
van der Klis \cite{hasinger89}), even though X-ray flux is not
monotonically related to $S_{\rm a}$ (van der Klis et al.  \cite{vdk90};
van der Klis \cite{vdk94apj}; Prins \& van der Klis \cite{prins97}).
Our proposed interpretation is that the observed changes in the
frequency of the kHz QPOs in LMXBs are driven by changes in $\dot M$
through the innermost part of the accretion disk, and that while the
observed X-ray flux is not a good indicator of this mass accretion rate,
the X-ray colors track it much better.  Possible reasons for this
include anisotropies in the emission, redistribution of some of the
radiation over unobserved energy ranges, and matter that flows onto the
neutron star in a more radial inflow, or away from it in a jet.

We have found that in 4U 1728--34 the frequency separation between the
two simultaneous kHz QPOs, $\Delta \nu$, is not consistent with being
equal to the frequency of the burst oscillations.  This is the second
source, after 4U 1636--53, for which $\Delta \nu$ does not match
$\nu_{\rm burst}$.  Furthermore, in 4U 1728--34 $\Delta \nu$ is not
constant, but it decreases as the frequency of the lower kHz QPO
increases.  This is the first source in which the three oscillations are
seen simultaneously, and in which the separation between the frequencies
of the kHz QPOs is observed to change.

It has been proposed that the upper kHz QPO at $\nu_{\rm upp}$
represents the Keplerian frequency of the accreting material in orbit
around the neutron star at some preferred radius (van der Klis et al.
\cite{vanderklis96}), while the lower QPO, at $\nu_{\rm low}$, is
produced by the beating of $\nu_{\rm upp}$ with the spin frequency of
the neutron star, $\nu_{\rm spin}$, which has been identified with
$\nu_{\rm burst}$ (or half that; Strohmayer et al.  \cite{strohmayer96};
Miller et al.  \cite{miller98}).  If this interpretation is correct,
$\Delta \nu$ equals $\nu_{\rm spin}$ and should remain constant.  The
neutron star spin can not change by $\sim 20 - 30$ \% on timescales of
days to months.

There are now four sources for which $\Delta \nu$ has been observed to
vary:  Sco X--1 (van der Klis et al.  \cite{vanderklisetal97a}), 4U
1608-52 (M\'endez et al.  \cite{mendez98c}, 4U 1735--44 (Ford et al.
\cite{ford1735}), and 4U 1728--34.  It has been argued that these
variations can be attributed to the near-Eddington mass accretion rate:
White \& Zhang (\cite{white97}) propose a 35\,\% expansion of the
neutron star photosphere with conservation of angular momentum, and F.
K.  Lamb (1996, private communication) suggests that the height of the
inner disk increases, and the different values of $\Delta \nu = \nu_{\rm
upp} - \nu_{\rm low}$ reflect different values of $\nu_{\rm upp}$ at
different heights in the disk.

Contrary to what we observe in 4U 1728--34, these explanations imply
that $\Delta \nu$ should approach $\nu_{\rm spin}$ when, at low $\dot
M$, the kHz QPOs move to lower frequencies.  Our results show that in 4U
1728--34, $\Delta \nu$ is still significantly below $\nu_{\rm burst}$ at
the lowest inferred $\dot M$ values.  This seems to rule out the simple
beat-frequency interpretation of the kHz QPOs in LMXBs, and some of the
modifications introduced to explain the results of a variable $\Delta
\nu$.

Recent ideas allowing the observed beat-frequency to deviate slightly
from the true one (F.  K.  Lamb 1998, private communication), as well as
proposals discarding one or more elements of the basic beat-frequency
picture, but still predicting definite relations between the observed
frequencies (e.g., Stella \& Vietri \cite{stella99}; Titarchuk, Lapidus,
\& Muslimov \cite{titarchuk98}) therefore merit careful consideration.

\acknowledgements

We are thanked to Will Zhang for for his comments that helped us to
improve the original manuscript.  This work was supported in part by the
Netherlands Organization for Scientific Research (NWO) under grant PGS
78-277, by the Netherlands Foundation for research in astronomy (ASTRON)
under grant 781-76-017, and by the Netherlands Research School for
Astronomy (NOVA).  MM is a fellow of the Consejo Nacional de
Investigaciones Cient\'{\i}ficas y T\'ecnicas de la Rep\'ublica
Argentina.  This research has made use of data obtained through the High
Energy Astrophysics Science Archive Research Center Online Service,
provided by the NASA/Goddard Space Flight Center.

\onecolumn

\clearpage

\clearpage

\clearpage

\clearpage

\end{document}